\documentclass[showpacs,aps,pre,floatfix]{revtex4}
\usepackage{graphicx} 

\begin{document}

%_______________ Title, authors, institutions _______________
\title{Simulations of driven and reconstituting lattice gases}

\author{M. D. Grynberg} 

\affiliation{Departamento de F\'{\i}sica, Universidad Nacional de  
La Plata,(1900) La Plata, Argentina}

%________ Abstract ____________________
\begin{abstract}
We discuss stationary aspects of a set of driven lattice gases in which 
hard-core particles with spatial extent, covering more than one lattice 
site, diffuse and reconstruct in one dimension under nearest-neighbor
interactions. As in the uncoupled case [\,M. Barma {\it et al.}, J. Phys. 
Condens. Matter {\bf 19}, 065112 (2007)\,], the dynamics of the phase
space breaks up into an exponentially large number of mutually 
disconnected sectors labeled by a non-local construct, the irreducible
string. Depending on whether the particle couplings are taken attractive
or repulsive, simulations in most of the studied sectors show that both
steady state currents and pair correlations behave quite differently at
low temperature regimes. For repulsive interactions an order-by-disorder
transition is suggested.
\end{abstract}

%________ Pacs Numbers ______________________
\pacs{05.40.-a, 02.50.-r, 05.70.Ln, 87.10.Hk}
%_____________________________________________
	
\maketitle

%____________________________
\section{Introduction}
%____________________________

Since a general theoretical framework for studying nonequilibrium phenomena yet
remains elusive, our understanding of the subject partly has to resort to studies of
specific and seemingly simple stochastic models. One of the most investigated ones
in that context is a driven lattice gas (DLG) involving hard-core particle diffusion
under an external bulk field which biases the particle flow along one of the lattice
axes. Introduced by Katz, Lebowitz and Spohn \cite{KLS} in part with the aim of
studying the physics of fast ionic conductors \cite{Valles}, it has triggered a great
deal of research for almost three decades \cite{Zia,Marro}. In the high field limit this
system describes an asymmetric simple  exclusion process (ASEP) \cite{Derrida,
Schutz} which already in one-dimension (1D) under suitable boundary conditions 
has encountered applications as diverse as protein synthesis \cite{Chou}, 
inhomogeneous interface growth \cite{Robin}, and vehicular traffic \cite{Popkov}. 
Notwithstanding the deceptive simplicity of most DLG versions, actually slight 
modifications of kinetic Ising models \cite{Kawa}, the constantly maintained bias 
results in a net dissipative current (if permitted by boundary conditions), so the 
emerging steady state (SS) distributions are nonequilibrium ones.

As part of the ongoing effort in this context, here we investigate numerically 
stationary aspects of DLG with extended objects which in turn can dissociate and 
reconstruct themselves in a 1D periodic lattice. Although less frequently studied, 
exclusion processes with spatially extended particles date back to the work of 
MacDonald, Gibbs, and Pipkin \cite{Mac}, who introduced this concept as a simple
setting to understand the dynamics of protein synthesis. In that terminology, each
lattice site denotes a codon on the messenger RNA, while large particles stand for
ribosomes which, covering several codons, move through them stepwise and 
thereby produce the protein \cite{Mac}. Subsequently, this and other issues related
to diffusion of extended objects were revisited in various studies \cite{Alcaraz,
Chou,Menon,Dhar,us,Gunter,Dong,Gupta}. 

In common with some of these latter, the processes here considered involve hard-core
composite particles, hereafter termed $k$-mers, which occupy $k$ consecutive 
locations and diffuse by one lattice site in the presence of both an external drive and
other fragments of length $l < k$. Besides, we include nearest-neighbor (NN) 
interactions between individual particles (hard-core monomers), and allow for 
$k$-mer dissociation \cite{Menon,Dhar,us} in the course of their casual 
encounters with the otherwise non-diffusing fragments, e.g. $\bullet \bullet\circ
\bullet\circ\,\rightleftharpoons\,\circ\bullet\bullet\bullet\circ\,\rightleftharpoons\,\circ
\bullet\circ\bullet\bullet$\,, say for dimers approaching a monomer. Thus, although 
the dynamics preserves the total number of $k$-mers, their indentities as well as 
those of the fragments collided are not retained, these being instead recomposed 
throughout the process. At finite drifts and couplings  the system evolves to a 
nontrivial SS measure characterized by a macroscopic current, a central quantity of
interest which in the terminology of protein synthesis corresponds to the stationary 
protein production rate \cite{Chou}. When monomers are uncoupled (beyond their
excluded volumes), all SS configurations are equally likely under periodic boundary
conditions (PBC) but the full dynamics and stationary correlations are quite involved 
\cite{us,Gupta}. 

Interestingly, and irrespective of the monomer couplings, ergodicity is strongly 
broken as a result of the presence of the aforementioned fragments. As these do not 
diffuse explicitly (but only through dissociation with $k$-mers), they break up the
phase space into mutually disjoint and dynamical invariant subspaces or `sectors'
whose number turns out to grow exponentially with the lattice size \cite{Menon,Dhar,
us,BD}. Thus, the SS current and correlations are not unique but rather vary from one 
sector to another, ultimately depending on the initial distribution of fragments. In 
that regard, to characterize the full partitioning of the phase space here we follow the
ideas given in Refs.\cite{Menon,Dhar,us} for the case of dimers, and introduce a set 
of fictitious extended particles whose order along a non-local construct\, -namely the 
`irreducible string' \cite{BD}-\, comes out to be the actual invariant of the motion.
This also enables us to obtain saturation currents of generic sectors by means of
a correspondence to ASEP systems \cite{us}. When normalized to those saturation 
values, our simulations indicate that as long as monomer interactions are held 
attractive the currents of most studied sectors  can be made to collapse into a single
universal curve. Moreover, this latter can also be fitted in terms of mean-field DLG 
currents \cite{Marro,Garrido} upon using the ASEP densities associated to each 
sector. By contrast, for repulsive interactions such normalized currents turn out to be
sector dependent and no universality can be constructed. On a mesoscopic level of 
description, also different features show up depending on whether the particle 
couplings are taken attractive or repulsive. Although in either case most sectors bear
highly degenerate ground states, based on the behavior of both the structure factor 
and correlation length at low temperature regimes, we suggest that for repulsive 
couplings thermal fluctuations appear to lift part of this degeneracy and cause an 
order-by-disorder transition \cite{Villain}.

The layout of this work is organized as follows. In Sec. II we define the basic kinetic
steps and transition probability rates of these reconstituting processes. We then
recast the dynamics in terms of new extended particles which readily evidence 
the appearance of an exponentially growing number of disconnected subspaces. Also,
for large drives these new particles are helpful to characterize the mentioned analogy
between reconstructing DLG and ASEP systems. Guided by these developments, 
simulations for dimers and trimers are discussed in Sec. III where we examine SS
currents and pair correlations in several sectors under various situations. We close
with Sec. IV which contains a recapitulation along with brief remarks on open
issues and possible extensions of this work.

%________________________________________________
\section{Diffusion of composite particles}
%________________________________________________

The microscopic particle model we consider is a ring of $L$ sites each of which may
be singly occupied (occupation number $n =1$), or empty ($n =0$). The particles
behave as if they were positive ions in relation to a uniform electric field $\cal E$,
while in turn are coupled effectively either by NN attractive interactions ($J > 0$), 
or NN repulsive ones ($J < 0$), via an Ising Hamiltonian $H = - 4 \,J \sum_i \, n_i \, 
n_{i+1}$. Let us first describe the basic kinetic steps which take place and then 
carry on with the definition of their corresponding rates. The system evolves 
stochastically under a particle conserving dynamics involving just $k$-mer shifts
in single lattice units, i.e. 
\begin{equation}
\underbrace{ 1 \dots 1}_k\, 0\;\; \rightleftharpoons \;\; 
0 \, \underbrace{ 1 \dots 1}_k\;,
\end{equation}
the motion being biased in the direction of the field. Here, monomers and groups or
fragments of $j$-adjacent particles with $j < k $ can not diffuse explicitly but since
the identity of  $k$-mers is impermanent, they are ultimately allowed to in a series 
of steps. For instance, in the sequence 
\begin{equation}
\label{reconstruction}
\underbrace{1 \dots 1}_k\, 0 \,\underbrace{1 . \, . 1}_j \,0 \;
\rightleftharpoons \;0\,  \underbrace{1 .\,. 1}_j \,
\underbrace{1 \dots 1}_k\,0 \; \rightleftharpoons \; 
0\, \underbrace{1 .\, . 1}_j\, 0 \,\underbrace{1 \dots 1}_k\,,
\end{equation}
the rightmost group of $j$ particles can hop $k$-sites to the left and vice versa. The 
key issue to emphasize is that both $k$-mers and fragments can dissociate and 
reconstitute without restrains throughout the process, so they do not retain their 
indentity (except in particular situations, as we shall see below).

As for transition rates between two particle configurations ${\cal C}$
and ${\cal C'}$, we take up the common Kawasaki transitions $\phi [ \,\beta
(\Delta H + u\, {\cal E} \,k)\,]$ with $\phi (x) \equiv 2 (1+e^x)^{-1}$, and 
$\Delta H = H ({\cal C'}) - H ({\cal C})$ \cite{Kawa,choices}. Here, the term
$u\, {\cal E}\, k$ denotes the work done by the bulk field $\cal E$ during a 
$k$-mer jump ($ u = \pm 1$), whereas $\beta = 1/T$ stands for the usual 
inverse temperature (henceforth the Boltzmann constant is set equal to 1). 
The different situations encompassed by these rules along with the rates 
associated to them are easily visualized in Table~\ref{tab1} (where thereafter 
$E_k \equiv \beta\, {\cal E}\, k$). In the absence of drive the SS distribution
is of course proportional to $e^{-H ({\cal C})/T}$ but owing to PBC, when 
${\cal E} \ne 0$ these rates no longer satisfy detailed balance \cite{Kampen}
and the exact form of the nonequilibrium SS distribution is unknown.
% (even for the monomer case).

%________________________ Table I __________________________________
\begin{table}[htbp]
\vskip 0.3cm
\begin{center}
\begin{tabular} { c  c } \hline
Process  &    \hskip 1cm Rate $( \rightleftharpoons)$ 
\\   \hline 
& \vspace{-0.3cm} \\

%_________________ 1st Process _____________________________________ 
$\circ \; \, \framebox{\phantom{------}}\, \circ \, \circ \;\; \;\, 
\rightleftharpoons \;\, \;\;  \circ \, \circ\,  \framebox{\phantom{------}} \; \, 
\circ$  &  \hskip 1cm  $\frac{1}{2} \left[\, 1 \pm \tanh \left( E_k / 2 \right)
\,\right]$
\vspace{0.3cm}\\

%_________________ 2nd Process _____________________________________ 
$\bullet \; \, \framebox{\phantom{------}}\, \circ \, \bullet \;\; \;\, 
\rightleftharpoons \;\, \;\; \bullet \, \circ \,  \framebox{\phantom{------}}
\; \, \bullet$ &  \hskip 1cm $\frac{1}{2} \left[\, 1 \pm \tanh \left( E_k / 2
\right) \,\right]$
\vspace{0.3cm}\\

%_________________ 3rd Process _____________________________________
$\bullet \; \, \framebox{\phantom{------}}\, \circ \, \circ \;\; \;\, 
\rightleftharpoons \;\, \;\; \bullet \, \circ \,  \framebox{\phantom{------}} \; \, 
\circ $ & \hskip 1cm  $\frac{1}{2} \left[\, 1 \pm  \tanh \left( E_k / 2 - 
2 \beta J  \right) \,\right]$
\vspace{0.3cm}\\

%_________________ 4th Process _____________________________________
$\circ \; \, \framebox{\phantom{------}}\, \circ \, \bullet \;\; \;\,  
\rightleftharpoons \;\, \;\; \circ \, \circ \,  \framebox{\phantom{------}} \; \, 
\bullet$ & \hskip 1cm $\frac{1}{2} \left[\, 1 \pm  \tanh \left( E_k / 2 +  
2 \beta J  \right) \,\right]$
\vspace{0.3cm}\\

\hline
\end{tabular}
\end{center}
\caption{Transition rates of driven $k$-mer processes in 1$d$. 
The symbols $\bullet\, , \, \circ,$ and\, \fbox{$^{\;\;\;\;\;\;\;\;\;}$} denote 
respectively occupied, vacant, and $k$-mer locations, whereas upper and 
lower signs stand in turn for forwards $\rightarrow$ and backwards 
$\leftarrow$ hopping rates.}
\label{tab1}
\end{table}
%___________________________________________________________________

This set of driven and reconstituting lattice gases (DRLG) may be viewed as an 
extended and interacting version of a dimer model recently studied in Ref.\,\cite{us}.
Alike its non-driven (equilibrium) predecessors \cite{Menon,Dhar}, the dynamic here 
considered splits up the phase space of configurations into an unusual large number
of invariant sectors, actually growing exponentially with the system size (see Sec. 
II\,B). Apart from particle conservation within each of the $k$ sublattices (say for 
integer $L/k$), the full partitioning of the phase space can be understood in terms of 
a nonlocal construct known as the irreducible string (IS) \cite{us,Menon,Dhar,BD}.
This latter is an invariant of the stochastic motion and in turn provides a convenient
label for each sector. To explain further this idea we recur to an equivalent 
representation of these processes using a set of composite
characters or new `particles' $A_0\,, A_1\,, \dots\,,A_{k-1}\,,A_k$ constructed as
\begin{eqnarray}
\nonumber
A_0 & \equiv &_{_{\hskip -0.39cm \vdots}}\;\;\, 0\,,\\
\label{species}
A_j & \equiv &_{_{\hskip -0.39cm \vdots}}\;\; \underbrace{ 1 \dots 1}_j\, 0\;, 
\;\; 1 \le  j <  k\,,\\
\nonumber
A_k & \equiv &  1 \dots 1\,.
\end{eqnarray}
The $A_k$ or $k$-mer movements and their recompositions can then be thought of
as character exchanges of the form
\begin{equation}
\label{exchange}
A_k \, A_j \;  \rightleftharpoons \; A_j \, A_k\;, \;\; 0 \le  j <  k\,,
\end{equation}
the $k$-mer identity here being preserved only by $A_0$, whereas exchanges not
involving $A_k$ remain disabled, i.e. $A_i \, A_j $ do not swap their positions if 
$i,j \ne k\,$. For example, in this representation the steps referred to in 
(\ref{reconstruction}) now become $A_k\,A_0\,A_j \,\rightleftharpoons\, A_0\,A_k\,A_j
\,\rightleftharpoons\,A_0\,A_j\,A_k \,$. This bears some resemblance to a driven 
process of several particles species introduced in the context of 1D phase separation
\cite{Evans}. However, in those systems all NN species are exchangeable 
\cite{Evans} whereas in this mapping note that the ordering of characters $A_0, \,
A_1, \, \cdots \,,A_{k-1}$ set by the initial conditions is {\it conserved} throughout 
all subsequent times, modulo eventual interposition of one or more adjacent 
$k$-mers between $A$'s. Thus, the invariant IS of a given sector simply refers to the 
sequence emerged after deletion of all $k$-mers or 'reducible' characters appearing 
in any configuration of that sector. This operation results in a unique sequence
irrespective of the order of deletion, so the correspondence between original
monomer and character configurations is one-to-one.

For this new representation we can therefore think of an equivalent Hamiltonian
$\cal H$ of hard-core particle species $A_i$ of fixed concentration $N_i/N$, defined
on a ring of $N = \sum_i N_i$\, sites with NN interactions. In terms of the occupation
numbers $n_{\lambda}^{(i)}$ of these new particle classes, the related Hamiltonian
can be written down as
\begin{equation}
\label{hamiltonian}
{\cal H} = -4 \,J\, \sum_{i \ne 0}\, \sum_{\lambda}\, n_{\lambda}^{(k)} 
n_{\lambda + 1}^{(i)}\;\,,\;\, \sum_{\lambda}\,n_{\lambda}^{(i)} = N_i\,,
\end{equation}
up to a sector dependent constant (here, if  $n_{\lambda}^{(i)} = 1$ at a given
location $\lambda$, then $n_{\lambda}^{(j)} = 0,\, \forall\, j \ne i\,$). Hence, the 
Kawasaki rates of the biased exchanges referred to in (\ref{exchange}) now depend 
on the class of surrounding particles, namely
\begin{eqnarray}
\nonumber
&\phantom{}_{\phi [\, \beta\, (\Delta {\cal H} - {\cal E} \,k)\,] }& \\
\label{A's}
A_i \, A_k\, A_j \, A_l \!\!\!\!\!\!\!\!\!\! & \rightleftharpoons &
\!\!\!\!\!\!\!\!\!\! A_i \, A_j \, A_k \,A_l\;\,,\;\, j \ne k\,,\\
\nonumber
&\phantom{}_{\phi [ -\beta\, (\Delta {\cal H} - {\cal E} \,k)\,] }&
\end{eqnarray}
with $\phi (x)$ defined as above,  and energy changes given by
\begin{equation}
\label{changes}
\Delta {\cal H}= \cases{0 \;\;\;\;\; {\rm if}\;\; l \ne 0\;\;\; 
{\rm or}\;\;\; l = j = 0\,,\, i \ne k\,,\cr
4 \,J\;\; {\rm otherwise}\,.}
\end{equation}
In particular, it follows that all IS sectors having $N_0 = 0$ {\it conserve} 
the internal energy throughout.

It is worth pointing out that for $k=1$ the form of 
Eqs.\,(\ref{hamiltonian}-\ref{changes}) just describes the usual monomer DLG. 
Also, notice that the dynamics of this latter is formally analogous to that of the
{\it non} reconstructing or null sector $[A_0]^{\cal L}$, that is, an IS of length
${\cal L} = N_0$. In that case Eq.\,(\ref{hamiltonian}) reduces to the usual Ising 
Hamiltonian, and the processes of (\ref{A's}) just involve $A_k$-$A_0$
('particle-hole') exchanges. In passing, we mention that it is actually the 
non-interacting version of this sector the one which was studied  in connection to 
the protein dynamics referred to in Sec. I, and the one whose space-time 
correlations were recently investigated in Ref.\,\cite{Gupta}.

%_______________________________
\subsection{The ASEP limit}
%_______________________________

For $N_0 = 0$ as well as in the large field or saturation regime 
$\vert {\cal E} \vert \gg \vert J \vert /k$ of generic sectors,  clearly the 
above processes are isomorphic to an ASEP system in which $k$-mers play the
role of {\it non-interacting} (but hard-core) particles hopping through 
indistinguishable $A_i$ vacancies ($i \ne k$). More specifically, given an IS of
length ${\cal L} = \sum_{i \ne k} (i+1)\, N_i$ measured in the original DRLG 
spacings, then this regime amounts to a problem of $N_k = (L - {\cal L})/k$ 
ASEP particles hopping with biased probabilities $\frac{1}{2}(1 \pm \tanh 
\frac{E_k}{2})$ through $N = \sum_i N_i$ lattice sites with particle density
\begin{equation}
\label{rasep}
\rho_{_{ASEP}}^{-1} = 1+ \frac{k}{L - \cal L}\,\sum_{i \ne k} N_i\,.
\end{equation}

From these correspondences we can readily obtain the sublattice currents of the
original DRLG in their saturation regimes. Since in our ASEP limit all SS configurations
are equally likely (because of PBC), evidently the probability of finding an $A_k$
particle followed by an $A_i$  vacancy or vice versa, is just $\rho_{_{ASEP}} \left( 1- 
\rho_{_{ASEP}} \right)\,$, in turn proportional to the ASEP current. In the DRLG
representation this is related to the probability of finding a $k$-mer 'head' ('tail') 
-\,i.e. a  rightmost (leftmost) $k$-mer unit\,- on a given sublattice site, times the 
probability of finding a $0$ in the next (previous) sublattice location. However, the 
former event occurs with probability $N_k/(L/k) = \rho_{_{ASEP}}\, k\,N/L\,$, whereas
that of the latter, $\propto \left( 1- \rho_{_{ASEP}} \right)$, must be normalized by 
the fraction of $0's$ of the sublattice in question. So if $\rho_i$ denotes the monomer
density of sublattice $\Lambda_i$, such fraction is therefore calculated as 
\begin{equation}
f_i = \frac{L}{k} (1 - \rho_i)\,/ \sum_{j \ne k} N_j\,,
\end{equation}
hence, for $\vert {\cal E} \vert  \gg \vert J \vert /k$ the saturation current of 
$\Lambda_i$ finally reduces to
\begin{equation}
\label{saturation}
 {\cal J}^{(i)}_{sat} = \rho_{_{ASEP}} \left( 1 -  \rho_i \right)\,
\tanh \left( E_k /2 \right)\,,
\end{equation}
while also holding $\forall\,{\cal E}$ so long as $N_0 = 0$. In general, these
saturation currents depend on the particular distribution of characters in the IS. 
However for periodic sectors, i.e. strings formed by repeating a unit sequence of 
$A's$, these limiting values are just rational functions of sublattice densities (cf. 
Table~\ref{tab2} below). In particular and for ulterior comparisons, in the null sector
$[A_0]^{\cal L}$ referred to above all sublattices share a common saturation current 
\begin{equation}
\label{saturation0}
{\cal J}_{sat} = \frac{\rho \,( 1 -  \rho )\,\tanh \left( E_k /2 \right)}
{\rho + k\, (1-\rho)}\,
\end{equation}
and a particle density $\rho = 1 - {\cal L}/L$.

Although in Sec. III we shall restrict ourselves to time independent SS aspects,
such as currents and pair correlations, let us briefly mention here that density
fluctuations in the stationary ASEP move through the system as kinematic
waves  \cite{Schutz,us,Gupta} that ultimately take over the asymptotic
behavior at large times. In our DRLG model this corresponds to $k$ sublattice
wave velocities $V_i = \partial {\cal J}^{(i)}_{sat} / \partial \rho_i$  which,
for periodic strings, can vanish at a common critical length ${\cal L}_c$. 
Hence, as suggested in Refs.\,\cite{us,Gupta}, when approaching such 
conditions there may well be a crossover from an exponential relaxation of 
density fluctuations to a slow Kardar-Parisi-Zhang dynamics \cite{Schutz}
for which the former would decay as $t^{-2/3}$. In particular, from 
Eq.\,(\ref{saturation0}) it follows that in the null sector this should occur for 
$\rho_c   \to  \sqrt k/(1+\sqrt k)$ \cite{Gupta}.

%_______________________________________________
\subsection{Growth of invariant sectors}
%_______________________________________________

Before proceeding to the simulation of these processes at finite temperatures and
fields, we pause to digress about the exponential growth of disjoint sectors with 
the length of their IS's. Here we follow and extend slightly the recursive procedure 
discussed in Ref.\,\cite{Menon} for the case of dimers. For simplicity, and solely
for the purpose of avoiding the overcount of strings related each other by a 
cyclic permutation of characters, we will assume open boundary conditions
throughout this subsection.  

Let $F_{\cal L} \, (1)$  and $F_{\cal L} \, (0)$ denote the number of IS's of 
length $\cal L $ whose first bit is 1 and 0 respectively. Thus, the  total number
of $\cal L$- sectors we want to evaluate is just $N_{\cal L} = F_{\cal L}\, (0) 
+ F_{\cal L}\, (1)$.  As there are no $A_k$ characters in these strings, then 
these quantities must satisfy the recursion relations
\begin{eqnarray}
\nonumber
F_{\cal L }\,(0) &=& F_{{\cal L} -1 }\,(0) + F_{{\cal L} -1 }\,(1)\,,\\
\label{recursion}
F_{\cal L }\,(1) &=&  F_{{\cal L} -1 }\,(0) + F_{{\cal L} -1 }\,(A_1) +
\dots + F_{{\cal L} -1 }\,(A_{k-2})\,,
\end{eqnarray}
with $F_{\cal L} \, (A_i)$  denoting in turn the number of  IS's of length $\cal L$
whose first character is $A_i$\,,\, $1 \le  i \le k-1$ (note that such construct would 
not be well-defined for PBC). On the other hand, by definition these latter quantities
should also be related recursively as
\begin{eqnarray}
\nonumber
F_{\cal L }\,(A_1) &=& F_{{\cal L} -1 }\,(0)\,,\\
\nonumber
F_{\cal L }\,(A_2) &=& F_{{\cal L} -1 }\,(A_1) =  F_{{\cal L} -2 }\,(0)\,,\\
\label{recursion2}
& \vdots & \\
\nonumber
F_{\cal L }\,(A_{k-1}) &=& F_{{\cal L} -1}\,(A_{k-2}) = \dots 
= F_{{\cal L} - (k-1)}\, (0)\,.
\end{eqnarray}
As a byproduct of these relations, it follows that $N_{\cal L} = 2 F_{\cal L}\, (0) - 
F_{{\cal L} - k}\, (0)$, thus it is sufficient to focus attention on $F_{\cal L}\, (0)$. 
Inserting the above equations in Eq.\,(\ref{recursion})  we readily obtain a linear 
recursion for this quantity, namely the Fibonacci-like relation
\begin{equation}
\label{fibo}
F_{\cal L}\, (0) = \sum_{i=1}^k F_{{\cal L} - i}\, (0)\;,\;\, {\cal L} > k\,,
\end{equation}
whose general solution is bound up to the  $z_1, \dots\,,z_k$ zeros of the 
polynomial \cite{Lando}
\begin{equation}
P_k (x) = \sum_{i=1}^k x^{i-1}\! - \,x^k\,.
\end{equation}
Thus, Eq.\,(\ref{fibo}) reduces to the exponential form  $F_{\cal L}\, (0) =
\sum_{i=1}^k b_i \, z_i^{\cal L}$ with  $b$-coefficients that are in turn 
evaluated by fitting linearly the $k$ boundary terms  $F_1 (0),\, \dots\,,
F_k (0)\,$, e.g. $F_1 (0) = F_2 (0) = 1$ for dimers, and $F_1 (0) = F_2 (0) =
\frac{1}{2}\,F_3 (0) = 1$ in the case of trimers.  Specifically, for these two
latter situations, which we shall discuss in detail for some sectors later on, 
it turns out that for large ${\cal L}$ (say comparable to $L$), the total 
number of $\cal L$-strings grows as
\begin{equation}
N_{\cal L} \propto \cases {\;\;\;\;\;\;\;\,
\left[\frac{1}{2}\left(1 + \sqrt 5\right) \right]^
{\cal L}  \simeq  1.618^{\cal L},\; {\rm for} \;\; k = 2\,, \cr\cr
\left[\frac{1}{3} \left(1 + v^+  + v^- \right) \right]^{\cal L} 
 \simeq  1.839^{\cal L},\; {\rm for} \;\; k = 3\,, }
\end{equation}
where $v{^\pm} =\left( 19 \pm 3\,\sqrt {33}\,\right)^{1/3}\,$.

From the above calculations, note that the number of non-jammed sectors put
together, i.e. the sum of all those with lengths ${\cal L} \in [1, L-1]$, increases
as fast as the number of sectors with ${\cal L} = L$. Since these latter can not
evolve any further, each constitutes a separate sector having only one configuration
(as opposed to non-jammed strings which, from the ASEP analogy, bear ${\sum_i N_i 
\choose \frac{L- \cal L}{k}}$ state configurations).

%_______________________________________
{\it Growth of sectors with $N_0 = 0$.--} 
%_______________________________________
Following this line of reasoning, it is straightforward to determine also the number of
IS conserving the internal energy throughout. That is the situation referred to after 
Eq.\,(\ref{changes}), where no $A_0$ characters appear in the IS. For this sector 
we now define $G_{\cal L}\,(1)$ and $G_{\cal L }\,(0)$ as the number of invariant 
$\cal L$- strings having {\it no consecutive} 0's, and whose first bit is 1 and 0 
respectively. Thus, the counting of strings constrained by $N_0 = 0$ requires the 
evaluation of $G_{\cal L }\,(1)$. Clearly, by construction these numbers involve the 
relations
\begin{eqnarray}
\nonumber
G_{\cal L }\,(0)  &=& G_{{\cal L} - 1 }\,(1)\,, \\
\label{G}
G_{\cal L }\,(1) &=&  G_{\cal L }\,(A_1) + \dots + G_{\cal L }\,(A_{k-1})\,,
\end{eqnarray}
where, as before, $G_{\cal L} (A_i)$ refers to the number of irreducible 
sectors, now subject to $N_0 = 0$, having $A_i$ as their first character
$(1 \le i \le k-1)$. Also, it can be readily verified that these latter numbers
are involved recursively in the {\it same} form as their $F$'s counterparts
in Eq.\,(\ref{recursion2}). When using those relations in Eq.\,(\ref{G}), the
following linear recurrence immediately emerges
\begin{equation}
\label{truncated}
G_{\cal L}\, (1) = \sum_{i=2}^k G_{{\cal L} - i}\, (1)\;,\;\, {\cal L} > k\,.
\end{equation}
The characteristic polynomials $P_k (x)$ associated to the generic term of
this sequence now distinguish the parity of $k$-mers,  namely \cite{Lando}
\begin{equation}
P_k (x) = \cases{ \sum_{i=0}^{k-2}\, x^i - x^k,\; {\rm for}\;\, k\;\,
{\rm odd}\,,\cr\cr
\sum_{i=0}^{\frac{k}{2} - 1}\, x^{2 i} - x^{k-1},\; {\rm for}\;\, k\;\,
{\rm even}\,,}
\end{equation}
which along with the boundary terms $G_1 (1), \dots\,,G_k (1)$ determine the
specific form of $G_{\cal L}\; \forall {\cal L} > k\,$. As expected, the roots of the
above polynomials only yield exponential growth for $k > 2$ (evidently for dimers
$G_{\cal L}(1) \equiv 1$, in correspondence with the sole $\left [A_1\right]^
{{\cal L}/2}$ configuration). In the limit ${\cal L} \to \infty$ these sectors grow 
progressively faster as $k$ increases but in all cases slower than the respective 
$F's$ of the unrestricted sectors [\,Eq.\,(\ref{fibo})\,], as they should. In particular, 
for trimers it turns out that 
\begin{equation}
G_{\cal L} (1) \propto \left( w^+ + w^- \right)^{\cal L} \simeq 
1.325^{\,\cal L},\;  k = 3\,, 
\end{equation}
where  $w^{\pm} = \left ( \frac{1}{2} \pm 
\frac{1}{6}\sqrt {\frac{23}{3}} \right)^{1/3}\!\!$. 

%________________________________________
\section{Numerical results}
%________________________________________

Armed with the ASEP correspondence discussed before,  we have conducted 
extensive simulations of SS currents and pair correlations in several subspaces for
both dimers and trimers. In all cases, we evolved indpendent configurations for each 
of the studied sectors. The corresponding initial conditions were prepared by random
deposition of  $(L - {\cal L})/k$ ASEP monomers, that is  $A_k$ particles ($k=2,\,3$),
on a ring of $N_0+N_1+ \cdots +N_k$ sites. Here, we distinguished between $k-1$ 
type of ASEP vacancies, so we tagged them in the same order as that appearing in 
the particular IS's considered, either periodic or not. Subsequently, each ASEP particle
was duplicated (triplicated for the case of trimers), by adding another particle (two 
particles) over one (two) extra adjacent location(s) specially created for that purpose, 
i.e. $1 \to 11$ ($1 \to 111$). In turn, the tagged vacancies were replaced accordingly 
by $j$ consecutive 1's followed by a 0, that is $0 \to 1 \cdots 1\,0$, if they referred to
$A_j$ characters, while keeping all $A_0$'s as 0's. This defines an efficient algorithm
to produce generic configurations in the chosen IS sector within the original DRLG 
representation of $L$ sites. Owing to PBC there is a small hindrance however, as 
eventually cyclic shifts in one site (one or two, for trimers) might be necessary to 
maintain invariable all sublattice densities in the generated DRLG configurations.
These latter were then updated with the stochastic rules summarized in 
Table~\ref{tab1}, using chains of $L = 1.2 \times 10^4$ sites evolving typically up
to $10^5$ simulation steps. Each of these ones involved $L$ update attempts at
random locations, after which the time scale was increased by one unit, i.e. $t \to 
t +1$, irrespective of these attempts being successful. 

The above algorithm enabled us to average measurements over nearly $5 \times 
10^5$  histories originated from independent sector configurations, thus reducing
significantly the scatter of our data. We considered three typical periodic situations
which were afterwards compared with the null string [\,analogous to the monomer
DLG, as already mentioned in Eqs.\,(\ref{hamiltonian}-\ref{changes})\,]. These are 
specified in Table~\ref{tab2} along with their sublattice saturation currents and
densities,  in turn arising from Eq.\,(\ref{saturation}) and simple stochiometric
considerations. We also examined random strings generated from the ASEP version
of these periodic sectors by swapping randomly through the lattice the order of their
irreducible characters, thus keeping all relative concentrations.
%_____________________ Table  II  _______________________________
\begin{table}[htbp]
\begin{center}
\begin{tabular} { c  c  c c} \hline
$k$ & \hskip 0.75cm IS sector  &   \hskip 0.75cm  ${\cal J}_{sat} /
 \tanh\, (\frac{E_k}{2})$  &   \hskip 0.75cm  Density
\\   \hline 
& \vspace{-0.15cm} \\

%-----------------Dimers,  Sector 1 ------------------------------
2 &  \hskip 0.75cm
$[A_{_1}^2 A_{_0}]^{{\cal L}/5}$  &   \hskip 0.75cm $\frac{(\rho - 1)\,
(5 \rho - 2)}{\rho - 4}$  &  \hskip 0.75cm $\rho =  1 - \frac{3 \cal L}{5 L}$
\vspace{0.3cm}\\

%-----------------Dimers,  Sector 2 ------------------------------
2 &  \hskip 0.75cm
$[A_{_1} A_{_0}]^{{\cal L}/3}$  &  \hskip 0.75cm  $\frac{(3 \rho - 1)\,
( \rho - 1)}{\rho - 3}$  &  \hskip 0.75cm $\rho =  1 - \frac{2 \cal L}{3 L}$
\vspace{0.3cm}\\

%----------------- Dimers, Sector 3 ------------------------------
2 &  \hskip 0.75cm
$[A_{_1} A_{_0}^2]^{{\cal L}/4}$  &   \hskip 0.75cm  $\cases{
\frac{(1-\rho_1)\,(1 - 2 \rho_1)}{\rho_1 - 2} \cr\cr \frac{2 \rho_2\, 
(1 - \rho_2)}{3-\rho_2}}$  &  \hskip 0.75cm $\cases{\rho_1 =  1 - 
\frac{\cal L}{2 L}\cr \vspace {-0.2cm} \cr \rho_2 =  1 - \frac{\cal L}{ L} }$
\vspace{0.3cm}\\

%----------------- Trimers, Sector 1 ------------------------------
3 &  \hskip 0.75cm
$[A_{_2} A_{_1}^2 A_{_0}]^{{\cal L}/8}$  &   \hskip 0.75cm 
$\frac{(\rho - 1)\,(2 \rho - 1)}{\rho - 2}$  &  \hskip 0.75cm $\rho =  
1 - \frac{\cal L}{2 L} $
\vspace{0.3cm}\\

%----------------- Trimers, Sector 2 ------------------------------
3 &  \hskip 0.75cm
$[A_{_2} A_{_1} A_{_0}]^{{\cal L}/6}$  &  \hskip 0.75cm  
$\cases{0 \cr\cr \frac{(1-\rho_2)\,(1 - 2 \rho_2)}{\rho_2 - 2} \cr\cr
\frac{2 \rho_3\, (1 - \rho_3)}{3-\rho_3}}$  &  \hskip 0.75cm 
$\cases{\rho_1 =  1 \cr \vspace {-0.2cm} \cr  \rho_2 =  1 - \frac{\cal L}{2 L}
 \cr \vspace {-0.2cm} \cr \rho_3 = 1 - \frac{\cal L}{L} }$
\vspace{0.3cm}\\

%----------------- Trimers, Sector 3 ------------------------------
3 &  \hskip 0.75cm
$[A_{_2} A_{_1} A_{_0}^3]^{{\cal L}/8}$  &  \hskip 0.75cm 
$\frac{(\rho - 1)\,(8 \rho - 3)}{7 \rho - 12}$  &  \hskip 0.75cm 
$\rho =  1 - \frac{5 \cal  L}{8 L} $
\vspace{0.3cm}\\
\hline
\end{tabular}
\end{center}
\vskip 0.3cm
\caption{Limiting sublattice currents and densities for dimers and trimers
($k = 2,3$) in periodic strings of length $\cal L$ considered in the simulations
below. These are formed by repeating, e.g.  $[A_1^2 A_0] = [(10)(10)(0)]$ 
${\cal L}/5$ times, etc. In some sectors these quantities are common to all 
sublattices.}
\label{tab2}
\end{table}

%_________________________________
\subsection{Currents}
%_________________________________

The measurement of SS sublattice currents in these sectors involved the monitoring 
of transient regimes which for most temperatures and fields decayed typically in 
$\sim 10^3$ steps. Then, we averaged all currents along two further decades during 
which no significant fluctuations were observed. As usual, the sublattice currents 
${\cal J}_i = \frac{k}{L \, \Delta t} \left \langle \, N^+_i - N^-_i  \,\right \rangle_{t, 
\,t + \Delta t}$ can be defined operationally using the total number of forwards 
$N^+_i$ and backwards $N^-_i$ particle jumps within sublattice $\Lambda_i$, and 
averaging over all event realizations during an interval $(t,\, t+\Delta t)$. However to
avoid any dependence on that latter lapse, particularly inconvenient to monitor early 
non-stationary stages, instead we measured instantaneous correlators of the form 
\begin{eqnarray}
\nonumber
{\cal J}_i (E_k, \beta J, t) &=&  \frac{k}{L}\,\sum_{j \in \Lambda_i} 
\big\langle \left( n_{j+1} \dots n_{j+k-1} \right) \left( R^+_j \,n_j \,
\bar n_{j+k} - R^-_j \, n_{j+k}\, \bar n_j \right) \big\rangle_t\,,\\
\label{current}
R_j^{\pm} &=& \frac{1}{2}  \pm \frac{1}{2}
\tanh \left[ \, \frac{E_k}{2} +
2 \beta J\,  \left( n_{j+k+1} - n_{j-1} \right)  \right] \,,
\end{eqnarray}
where $\bar n_j \equiv 1 - n_j\,$ are just vacancy occupations numbers in 
sublattice $\Lambda_i$. Here, $\langle\, \rangle_t$ denotes an ensemble 
average over these correlators at time $t$, whereas right (left) hoppings
$R^+$ ($R^-$) are defined so as to take into account the rates referred 
to in Table~\ref{tab1}. 
%_______________ Fig.1 (currents vs. E) ____________________
\begin{figure}[htbp]
\vskip -1.2cm
\hskip -1.2cm 
\includegraphics[width=0.55\textwidth]{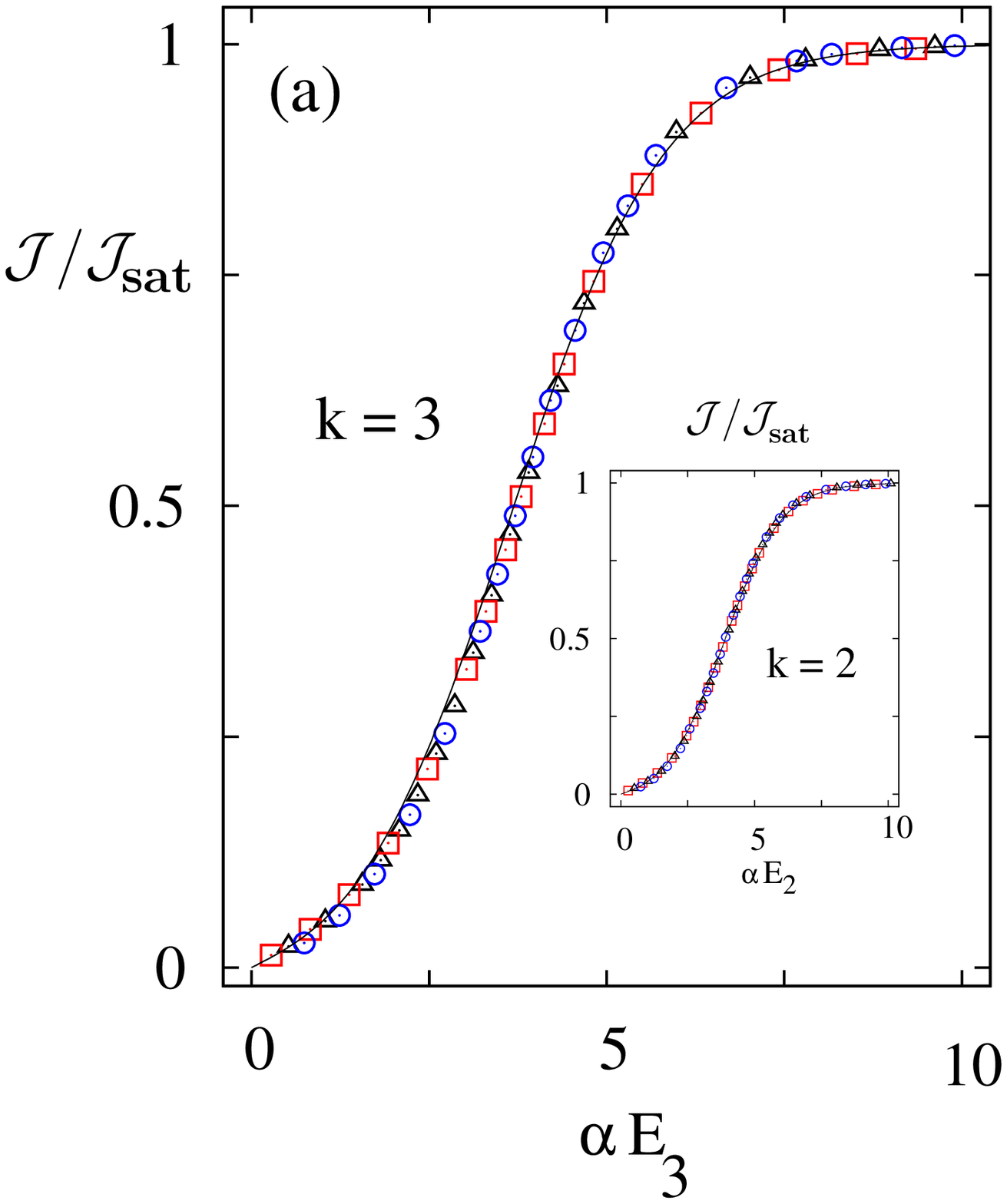}
\hskip -1.5cm
\includegraphics[width=0.55\textwidth]{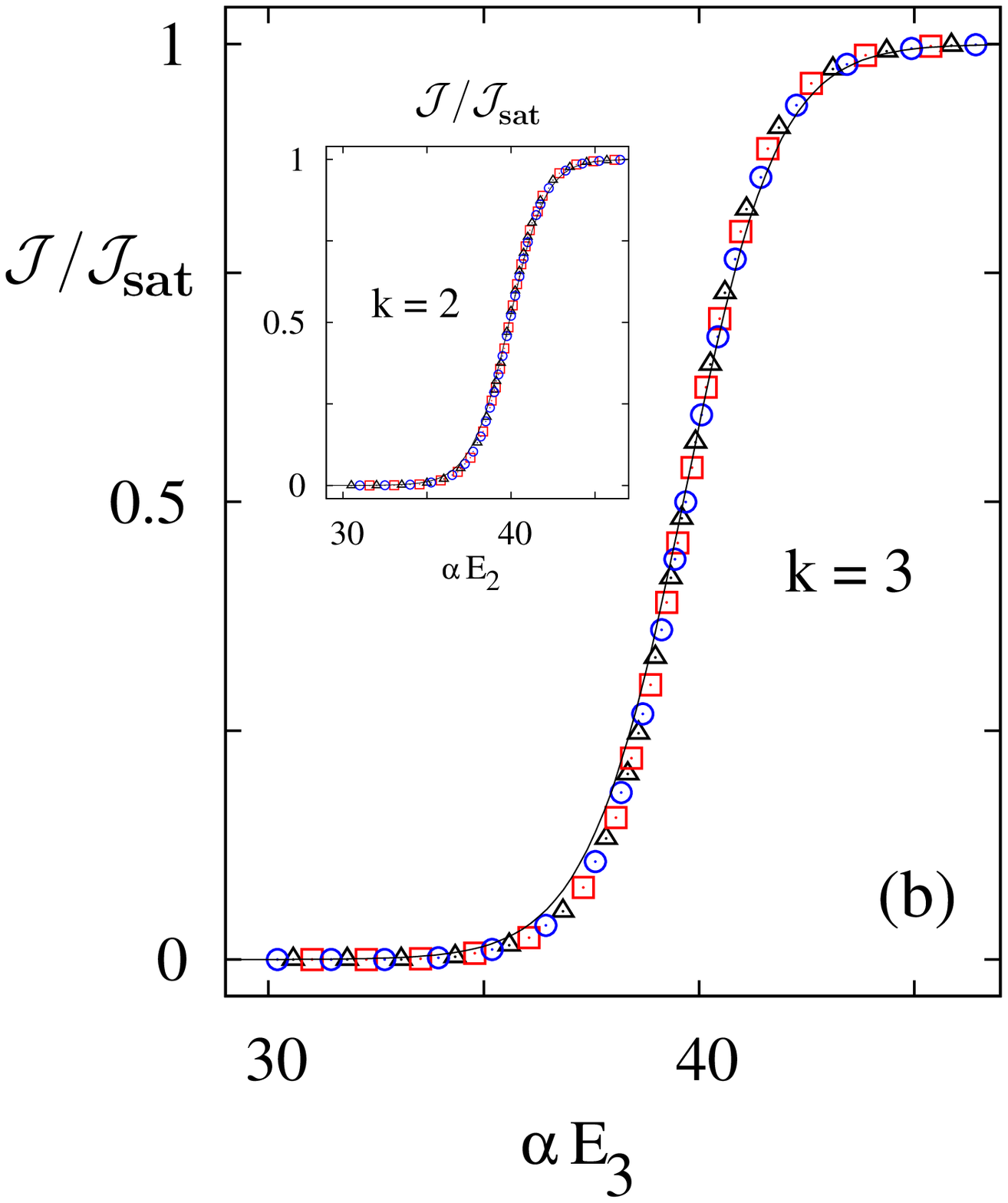}
\vskip -5cm 
\hskip -1.2cm
\includegraphics[width=0.55\textwidth]{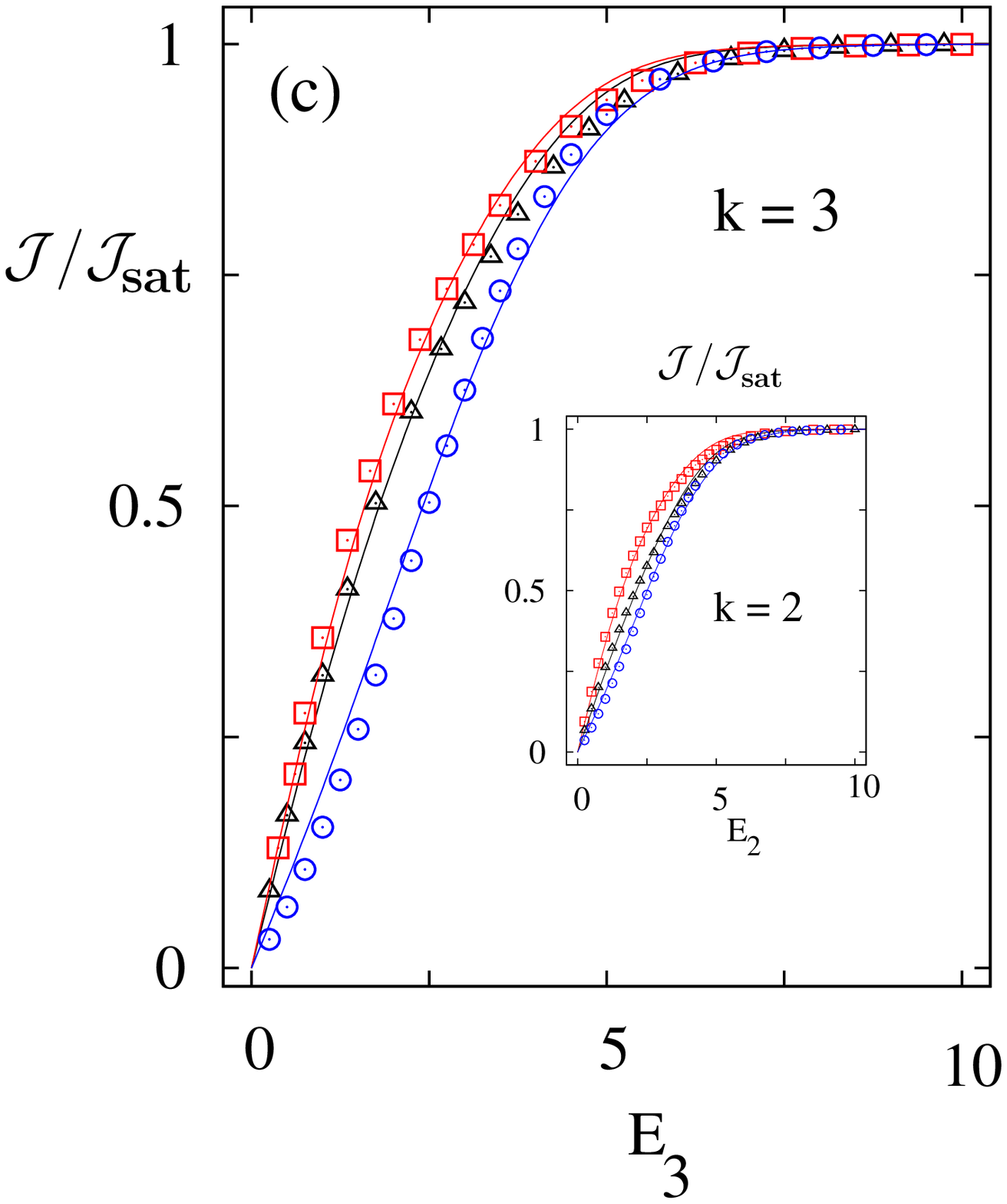}
\hskip -1.5cm
\includegraphics[width=0.55\textwidth]{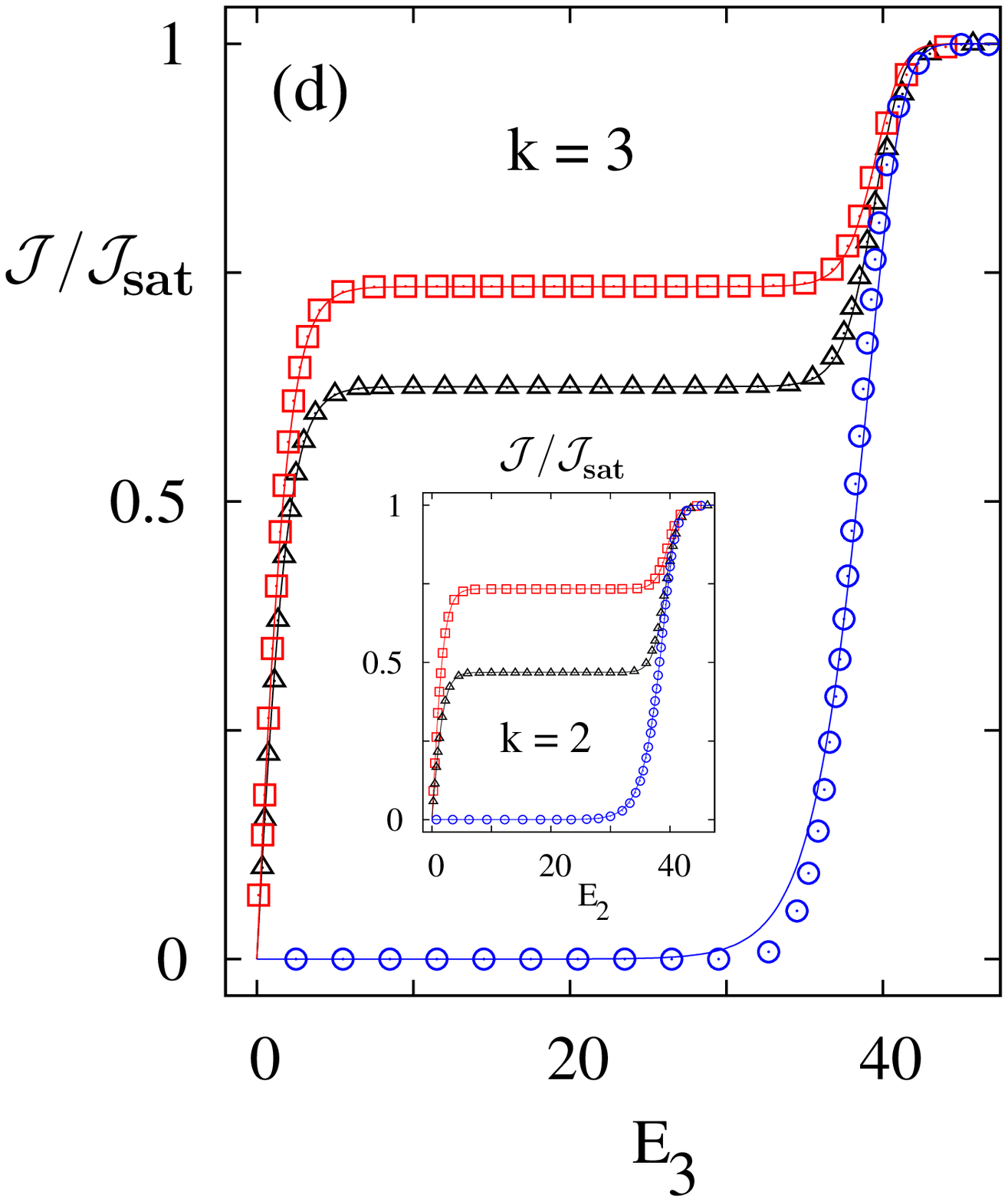}
\vskip -2.5cm
\caption{(Color online) Normalized SS currents for reconstructing trimers and dimers
(insets) using ${\cal L}/L= 1/2$. In listed order, the sectors of Table~\ref{tab2} here 
correspond to squares, triangles, and circles respectively. In (a) $T/J = 1$, and (b) 
$T/J = 0.1$, all data collapse onto the null string current (for clarity not shown), upon
rescaling the drifts $E_k$ with $\alpha$'s such that $\vert 1 - \alpha \vert \alt 0.05$. 
The data are fitted by the mean field currents of Refs.\,\cite{Garrido,Marro} (solid 
lines), using the associated  DLG densities referred to in Eq.\,(\ref{rasep}). The mean
field description extends as well to (c) $T/J = -1$, and (d) $T/J = -0.1$, where however 
all currents are sector dependent and DLG densities are chosen only as fitting 
parameters.}
\label{currentE}
\end{figure}
%_______________________________________________________
In Fig.\,\ref{currentE} we show the resulting SS currents normalized to the saturation
values of Table~\ref{tab2}, after taking $T/J = \pm 1,\,  \pm 0.1$, and ${\cal L}/L = 
1/2$ for several driving fields. In addition, Fig.\,\ref{currentT} displays other plane of 
the current, in this case holding $E_k = 10$ for a variety of temperatures, using both
attractive and repulsive couplings. \,\, It turns out that normalized currents of 
nonequivalent sublattices are indistinguishable within our error margins, a non 
obvious feature (except for random strings, as their sublattice occupations approach 
each other as ${\cal L} \to \infty$). 

{\it (i) $J > 0$}.--
More importantly, under attractive interactions the currents of all studied sectors, 
both periodic and random, can be made to collapse into that of the null string by 
rescaling slightly the driving fields. This is displayed in Figs.\,\ref{currentE}a and 
\ref{currentE}b (random sectors not shown, for legibility). Furthermore, the data
collapse extends also to the ${\cal J}-T$ plane provided the attractive couplings are 
taken slimly rescaled, as illustrated in Fig.\,\ref{currentT}a. It is noteworthy that in
both ${\cal J}-E_k$ and ${\cal J}-T$ planes the null sector data follow very closely the
mean field currents of the usual DLG \cite{Garrido}. These arise essentially from a 
kinetic version of the cluster variation method applied to dynamics proceeding via 
exchange processes \cite{Marro}. As expected from the arguments given in Sec. II\,A,
here the fitting of the null string current is attained upon choosing $\rho_{\!_{\cal L}}
= \left(1 - \frac{\cal L}{L} \right) / \left[ 1 +  \frac{\cal L}{L}(k-1)\right]$ as the 
monomer density for the DLG system [see Eq.\,(\ref{rasep})\,]. These numerical 
observations naturally lead us to put forward the universality hypothesis 
\begin{equation}
\label{universality}
{\cal J}^{(i)}_{_{DRLG}}  (E_k, \, \beta J,\,{\cal L}) / {\cal J}^{(i)}_{sat} = 
\frac{{\cal J}_{_{DLG}} (f \,E_k, \, g\, \beta J,\, \rho_{\!_{\cal L}}) }{\rho_{\!_
{\cal L}} \left( 1 -  \rho_{\!_{\cal L}} \right)}\;,\; J > 0\,,
\end{equation}
for normalized sublattice currents in ferromagnetic DRLG . Here, $f=f(\beta J)$ and 
$g=g( E_k)$ are sector dependent scaling factors (probably close to 1), whereas 
${\cal J}^{(i)}_{sat}$ is taken as in Eq.\,(\ref{saturation}). Preliminary runs using
other string lengths indicate similar results, thus adding more weight to this 
conjecture. 
%__________________ Fig. 2 (Currents vs. T / | J | ) __________________
\begin{figure}[htbp]
\vskip -1.35cm
\hskip -1.2cm 
\includegraphics[width=0.55\textwidth]{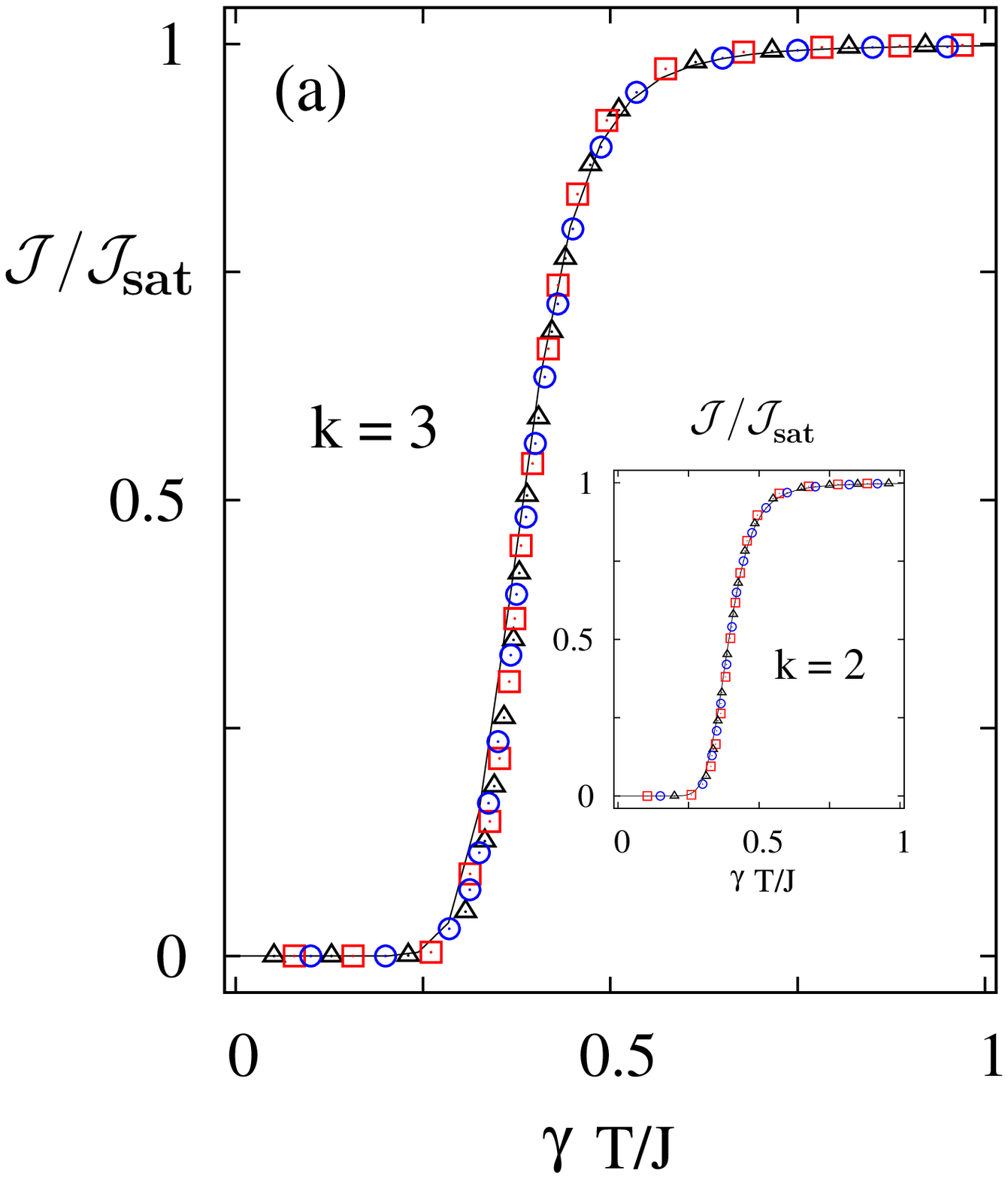}
\hskip -1.5cm
\includegraphics[width=0.55\textwidth]{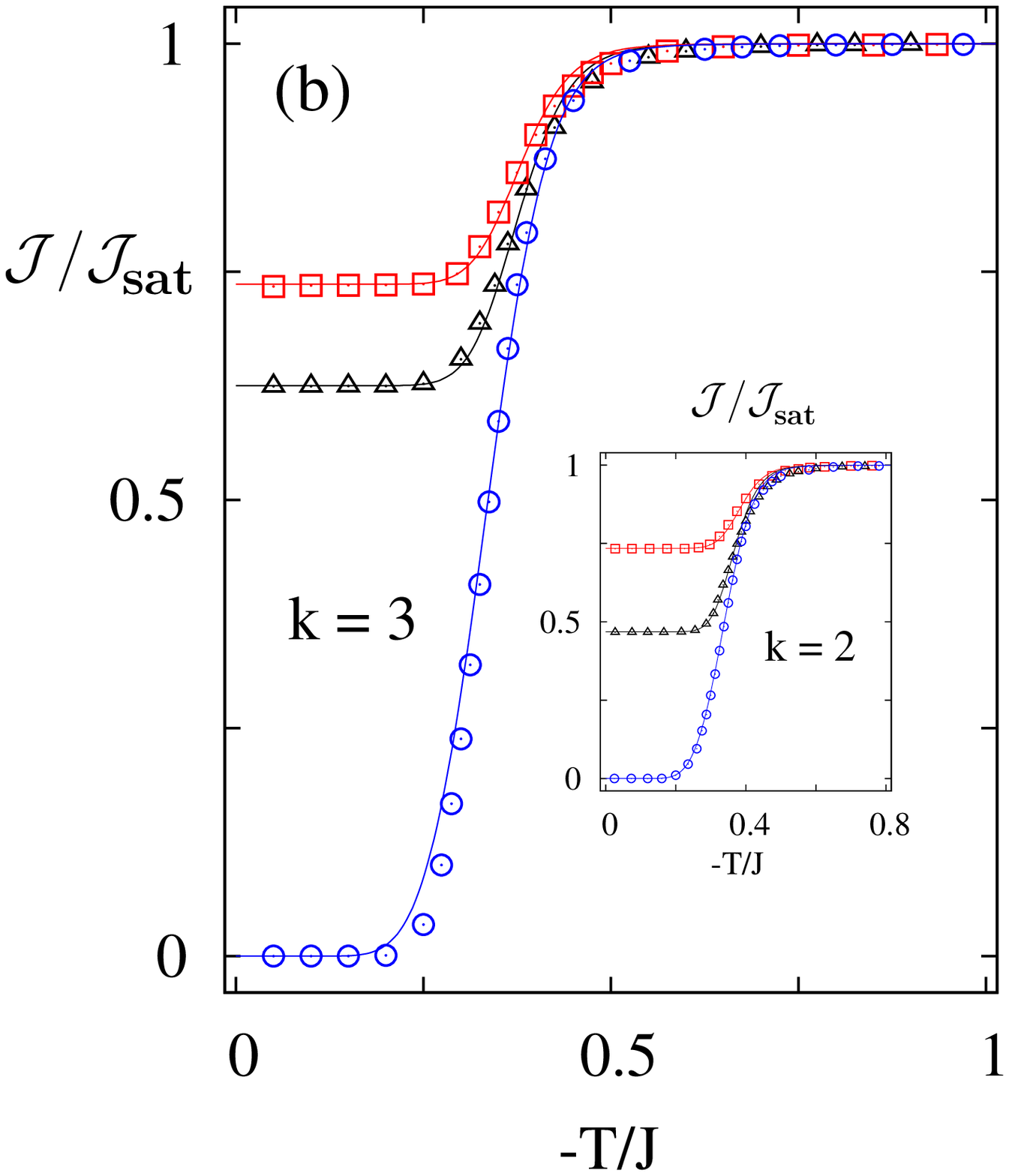}
\vskip -2.5cm
\caption{(Color online) Normalized SS currents versus temperature for $k=3$, 
and $k=2$ (insets) while keeping ${\cal E} k /T = 10$ for (a) $J > 0$,
and (b) $J < 0$.  Key to symbols is the same as in Fig.\,\ref{currentE}. In (a) 
the data collapse is obtained by the rescale $J \to J/\gamma,$ choosing 
$\gamma$'s such that $\vert 1 - \gamma \vert \alt 0.02$. These currents 
follow closely the null string data (not displayed to avoid overcrowding). 
By contrast, in (b) all currents are sector dependent. In both cases the mean
field fittings (Refs.\,\cite{Garrido,Marro}, solid lines) arise from the 
considerations given for Fig.\,\ref{currentE}.} 
\label{currentT}
\end{figure}
%______________________________________________________

{\it (ii) $J < 0$}.-- On the other hand, for repulsive interactions all normalized
currents come out to be sector dependent, as is shown in Figs.\,\ref{currentE}c, 
\ref{currentE}d, and \ref{currentT}b, an aspect becoming more pronounced as 
temperature is lowered. However, this dependence appears to involve only the 
relative concentration of irreducible characters rather than their particular 
distributions, because in all situations the normalized currents of random sectors
follow closely those of their periodic counterparts (alike the attractive case). Although
these currents can also be fitted using the mean field approach to DLG \cite{Garrido,
Marro}, the corresponding monomer densities can no longer be understood in terms 
of the ASEP analogy given before (except for null sectors and/or high drive regimes, 
as expected). Here, we merely use DLG densities as fitting parameters which actually
turn out to be rather sensitive in tuning the values of the current plateaux exhibited
in Figs.\,\ref{currentE}d, and \ref{currentT}b.

Further to universality issues, or the lack thereof for $J < 0$, these results also 
suggest that DRLG currents inherit some of the salient features of DLG ones according
to the type of interactions  \cite{Garrido,Marro} (c.f. nevertheless pair correlations in
non null sectors). For $J > 0$ there is a continuous changeover from a rather
insulating state at low temperatures and  fields to a conducting phase as both $T$
and $\cal E$ are increased. For $J < 0$ however, already at low temperatures some
sectors can be found in conducting states, while exhibiting comparatively larger 
conductivities at small fields. As for the appearance of current plateaux in 
Figs.\,\ref{currentE}d, and \ref{currentT}b, notice that these are already present at
the mean field level of the standard DLG.

%_________________________________
\subsection{Pair correlations}
%_________________________________

\begin{figure}[htbp]
\vskip - 2cm
\hskip -0.5cm 
\includegraphics[width=0.55\textwidth]{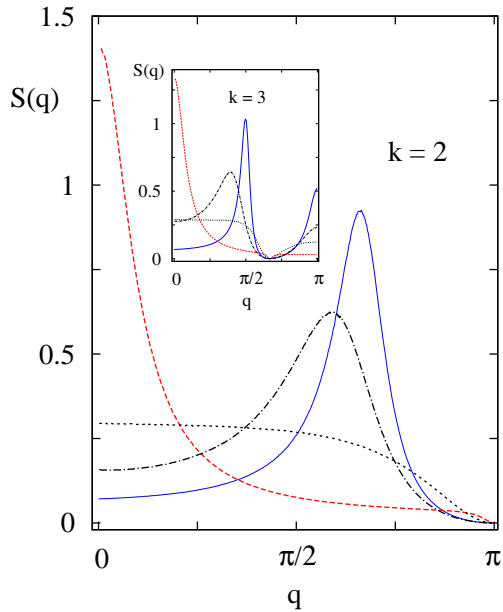}
\vskip -2.6cm
\caption{(Color online) Structure factors of sector $[A_0]^{\cal L}$ (null string) 
for dimers and trimers (inset), using ${\cal L}/L =$ 1/3 and 1/4 respectively. Solid,
dashed, and dotted lines stand in turn for $(E_k, T/J) = (2,\,-1), (2,\,1)$, and $(8,\,
\pm 1)$. The two first cases exhibit maxima very near to $q = 2 \pi/3 $ (dimers), 
$\pi / 2$ (trimers), and 0 (both), thus resembling the average equilibrium periodicity 
of low temperature regimes. At high fields no periodicity takes over, regardless of 
the sign of $J$. The dashed-dotted lines denote the situation $(E_k, T/J) = (2,\,-1/2)$ 
for ${\cal L}/L =$ 1/2, where the ground state is highly degenerate.}
\label{SFnull}
\end{figure}

Turning to mesoscopic scales, in the following we focus on the SS instantaneous
density-density correlation functions expressed through the subtracted or 
cummulant form 
\begin{equation}
C (r) = \frac{1}{L}\,\sum_j \big(\, \langle \,n_j\,n_{j+r} \,\rangle  -
\rho_{\Lambda_j}\, \rho_{\Lambda_{j+r}}  \big)\,,
\end{equation}
with $\rho_{\Lambda_i}$ being the density of sublattice $\Lambda_i$. Also, to gain
some further insight into the average organization of stationary regimes, we consider
the static structure factor or Fourier transform of $C (r)$
\begin{equation}
S (q) = \frac{1}{L}\! \sum_{-L/2 \, \le\,r \,<\,L/2} \!\!\!\!\!\!\!\!\!e^{i\,q\,r}\,C (r)\,,
\end{equation}
which in our case is a real function of the wavelength $\lambda = 2 \pi /q$. In 
Fig.\,\ref{SFnull} we first show this latter function for the case of dimers and trimers
in the null sector, taking ${\cal L}/L = 1/3$ and 1/4 in turn. At finite temperatures 
and far from the saturation or ASEP regime  $\vert {\cal E} \vert \gg \vert J \vert$, the
main maxima might be regarded as remnants of the periodicity and long range order
of the Ising ground states. Clearly, in the attractive and repulsive situations these 
are respectively of the form $A_k^{ L  /(k+1)} A_0^{L/(k+1)}$ and $[A_k\,A_0]^
{L/(k+1)}$ (along with translations), so as $T \to 0$ it is natural to expect sharp
peaks at $q = 0$ and $\frac{2 \pi}{k+1}$ in each case. However, in general notice
that when ${\cal L}/L \ne \frac{1}{k+1}$ the ground state of the null sector is highly
degenerate for $J < 0$ [\,having the form $A_k^{m_1} A_0\,A_k^{m_2}A_0  \cdots$
with $m_j \ge 0$\, constrained as $\sum_j m_j = (L - {\cal L})/k$\,], thus setting a 
residual entropy which grows linearly with the system size. In fact, as temperature is 
lowered the structure factor remains essentially broad (see data of $T/J = -1/2$ for 
${\cal L}/L = 1/2$), and correlation lengths are of the order of the lattice spacing. 
A similar scenario arises in the monomer DLG, where degeneracies for $J < 0$ and 
$\rho \ne 1/2$ also preclude long range order at any temperature \cite{Garrido,
Marro}.
%_______________Fig. 3 (Structure factors in the null sector) _________________
%__________________________________________________________________________

\begin{figure}[htbp]
\vskip - 1cm
\hskip -1.2cm 
\includegraphics[width=0.55\textwidth]{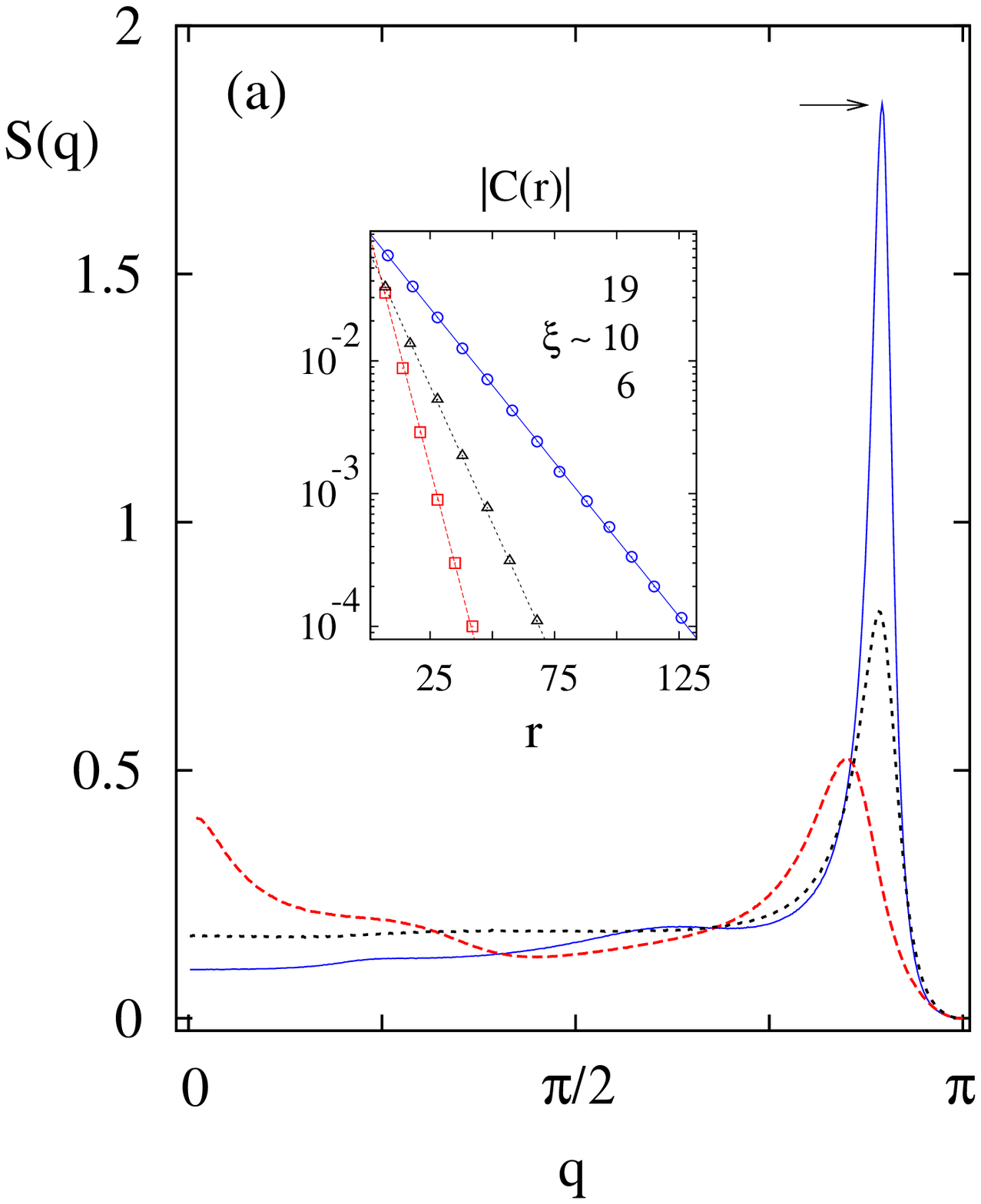}
\hskip -1.5cm
\includegraphics[width=0.55\textwidth]{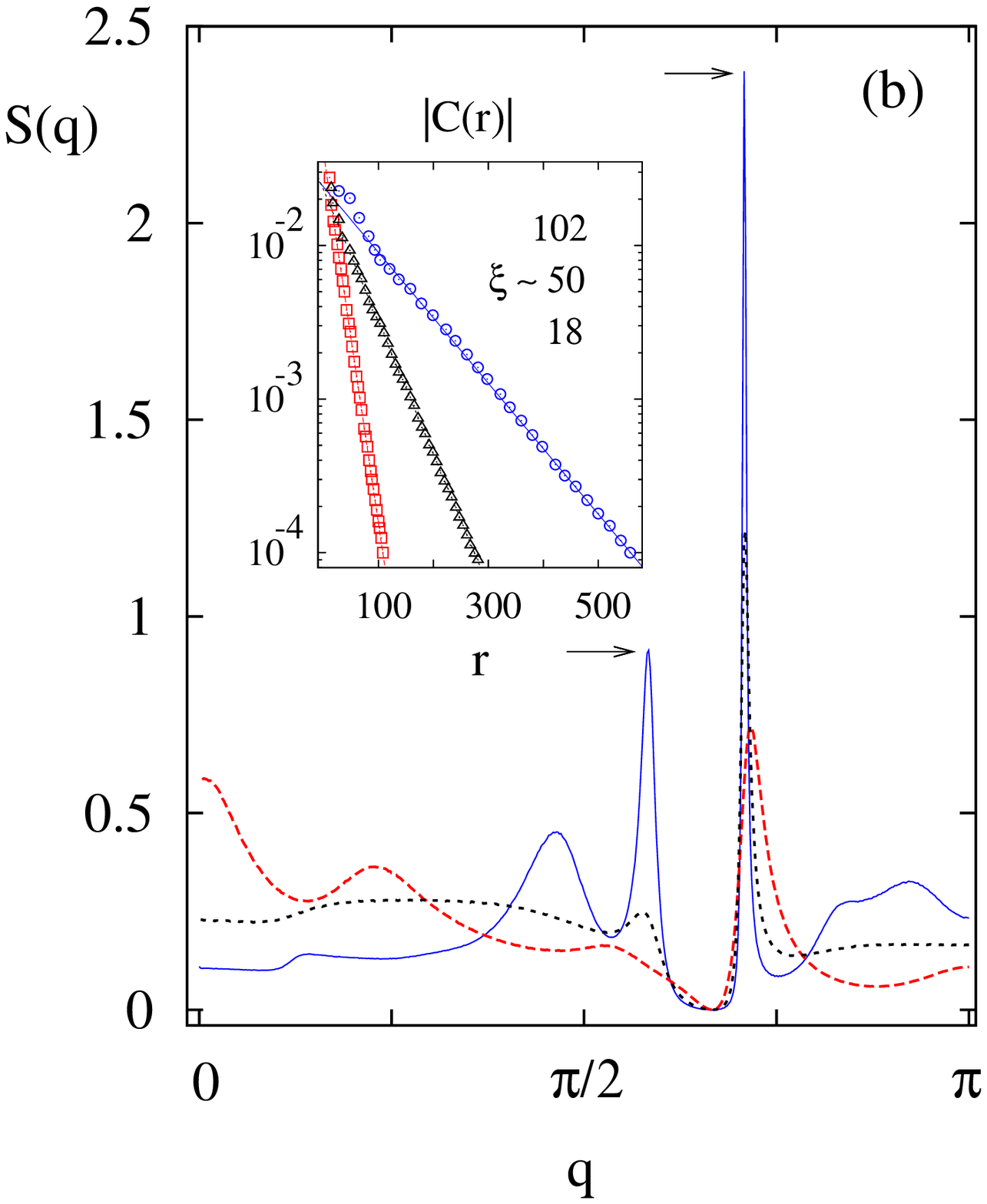}
\vskip -4.75cm
\hskip -0.25cm
\includegraphics[width=0.55\textwidth]{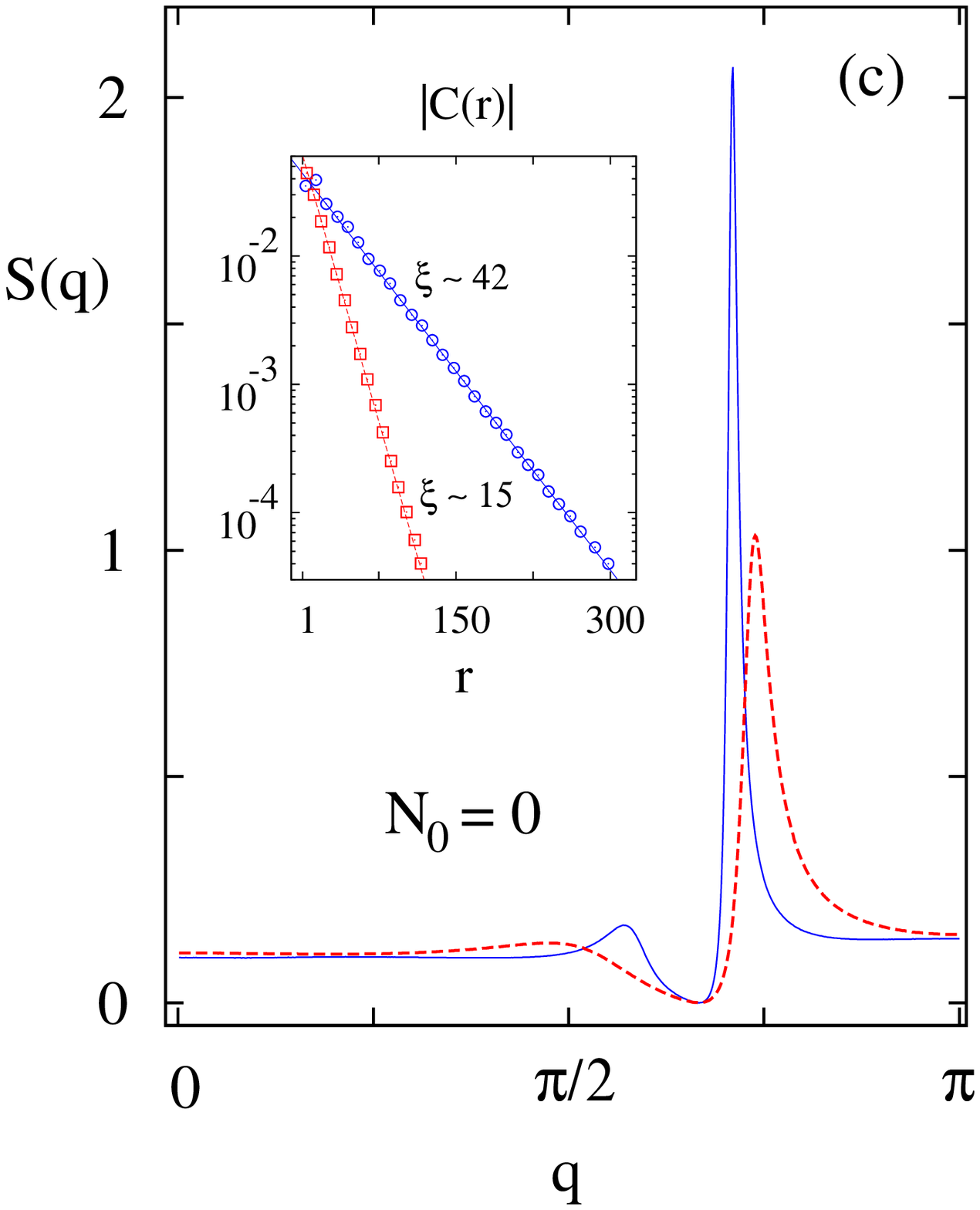}
\vskip -2.5cm
\caption{(Color online) Stationary structure factors for (a) dimers in sector 
$[A_{_1}^2 A_{_0}]^{{\cal L}/5},$ (b) trimers in  $[A_{_2} A_{_1} A_{_0}^3]
^{{\cal L}/8},$ (c) trimers in energy conserving sectors  $[A_{_2}^2 A_{_1}]^
{{\cal L}/8}$ (solid line, circles), and  $[A_{_2} A_{_1}]^{{\cal L}/5}$ (dashed line,
squares), using ${\cal L}/L = 1/2$ throughout. Spatial correlation functions (insets)
are highly oscillating, but only local maxima evidencing a decay $\propto e^{-r/\xi}$
are shown. Although (a) and (b) bear both highly degenerate ground states
irrespective of the coupling sign, for $J < 0$ the peaks marked by arrows rapidly 
grow as temperature is lowered. As in Fig.\,\ref{SFnull}, here solid, dashed, and 
dotted lines (in turn, circles, squares, and triangles for insets) denote respectively 
the cases $(E_k, T/J) = (2,\,-1), (2,\,1)$, and $(8,\,\pm 1)$. In the high field regime 
the results of (a) and (b) become independent of $T/J$, whereas in (c) this is always 
the case whether or not the system is driven.}
\label{SF}
\end{figure}

{\it (i) $J < 0$}.-- Yet, a rather different situation shows up for repulsive couplings in 
other periodic sectors also bearing high degeneracies. This is observed in 
Figs.\,\ref{SF}a and \ref{SF}b where, as an example, we illustrate respectively the 
behavior of strings $[A_1^2\,A_0]^{{\cal L}/5}$ ($k=2$), and $[A_2\,A_1 A_0^3\,]
^{{\cal L}/8}$ ($k=3$). 
%_____________Fig. 4 (Structure factors for dimers and trimers) ______________
%______________________________________________________
It can be readily verified that for $J < 0$ the number of ground states of the former 
grows exponentially with the lattice size so long as ${\cal L}/ L \ne 5/7$, whereas that
of the latter grows in the same manner provided ${\cal L}/ L \ne 8/17$\, (otherwise 
these states would be plain periodic sequences of the form $[\,A_1^2 A_2\,A_0 ]^
{L/7}$ for dimers, and $[\,A_2\,A_1\,(A_3\,A_0)^3\,]^{L/17}$ for trimers). However, 
despite that exponential degeneracy the structure factors of both sectors can single 
out wavelengths exhibiting peaks that rapidly narrow and heighten as temperature is 
lowered. Also in lowering $T$, correlation lengths turn out to grow monotonically (e.g.
at $T/J = -1/2$, $\xi \sim$ 500 in sector $[A_2\,A_1\,A_0^3\,]^{{\cal L}/8}$), while 
being already relatively large  at $T/J = -1$ as compared with those of the case $J > 0$
(see below). This is suggestive of an order-by-disorder scenario \cite{Villain} in which
thermal fluctuations are able to lift part of the degeneracy by selecting a subset of 
states with largest entropy in the ground state manifold. Here, note that the role of 
frustration in Refs.\,\cite{Villain} is being played by the string conservation laws and
the spatial extent of characters.

{\it (ii) $J > 0$}.--  By contrast, under attractive interactions no ordering seems to 
emerge for these sectors at low temperature regimes. Structure factors now remain 
basically flat (i.e. bounded) and correlation lengths do not increase as $T$ is lowered
(Figs.\,\ref{SF}a and \ref{SF}b). Thus, thermal fluctuations now appear as being 
unable to suppress the residual entropy (also growing linearly in the thermodynamic
limit of both sectors $\,\forall\,\, {\cal L}/ L \ne 0^+, 1^-)$ and no order-by-disorder 
seems likely to occur, just as in the situation of the null string and standard DLG 
under repulsive couplings. Similar results for $J > 0$ were observed in the other 
sectors of Table~\ref{tab2}, instead exhibiting effects of order by thermal 
fluctuations as long as the couplings are taken repulsive.
 
{\it (iii) $\vert {\cal E} \vert \gg \vert J \vert$}.-- In approaching the ASEP or high field
limit, pair correlations become progressively independent of the coupling values
because all configurations tend to be equally likely. Already for $(E_k, T/J) = (8, 
\pm 1)$, Figs.\,\ref{SFnull}, \ref{SF}a, and \ref{SF}b indicate that all results are 
numerically indistinguishable. However as the field is lowered from the ASEP regime, 
for $J < 0$ correlations can enhance significantly their ranges, in turn becoming 
arbitrarily long if temperatures are taken low enough. As displayed in Figs.\,\ref{SF}a 
and \ref{SF}b, this is to be contrasted with the opposite trend of both $\xi$ and $S(q)$
under attractive couplings, where the drive favors a slight increase of pair correlations.

{\it (iv) $N_0 = 0$}.--  Finally, the ASEP regime is also related to the energy 
conserving sectors ($N_0 = 0$) referred to after Eq.\,(\ref{changes}), for which the 
presence of NN interactions is inconsequential. Despite the fact that every state is 
equally weighted, owing to the $k$-mer size here structure factors can still single 
out characteristic wave numbers (sector dependent) and exhibit large correlation 
lengths. This is illustrated in Fig.\,\ref{SF}c. In the ASEP limit of the null sector these
issues were recently analyzed in Ref.\,\cite{Gupta} where closed expressions for
general $k$'s were given for static correlations. Such rich behavior contrasts to that 
of totally uncorrelated monomers and dimers in sectors $[A_1]^{{\cal L}/2}$ (the 
only ones with $N_0 = 0$, which can be viewed as hopping monomers within one 
independent sublattice).

%____________________________
\section{Concluding remarks}
%____________________________

To summarize, we have discussed stationary aspects of driven lattice gases in which
the role of biased monomers is played by extended and reconstructing $k$-mers 
under a Kawasaki dynamics [\,Eq.\,(\ref{reconstruction}) and Table~\ref{tab1}\,]. 
Exploiting the correspondence between these processes and those involving the 
particle species defined in Eq.\,(\ref{species}) we readily identified the many sector 
decomposition of the original problem, ultimately encompassed in the invariant 
ordering of these characters along the so called irreducible strings \cite{Menon,BD}. 
In the high field regime the dynamics of these new particles was thought of as an
asymmetric exclusion process defined on a smaller lattice (Sec. II\,A), thereby 
enabling us to evaluate saturation currents [\,Eq.\,(\ref{saturation})\,] for generic
strings or sectors of motion. In turn, the proliferation of these latter was shown to 
grow exponentially with the length of these strings (Sec. II\,B). 

At finite temperatures and drives, we studied numerically the case of both dimers 
and trimers in typical sectors whose initial configurations were prepared by random 
sequential adsorption of `monomers' in the equivalent ASEP states. These were then 
raised up into DRLG configurations, always keeping the distribution of irreducible 
characters or tagged ASEP `vacancies' in each of the studied sectors
(Table~\ref{tab2}). The emerging stationary currents clearly discern between 
universal (Figs.\,\ref{currentE}a, \ref{currentE}b, \ref{currentT}a), and sector 
dependent behavior (Figs.\,\ref{currentE}c, \ref{currentE}d, \ref{currentT}b) 
according to the particle couplings being attractive or repulsive. In the former case, 
the universality hypothesis put forward in Eq.\,(\ref{universality}) suggests in turn 
an effective medium relation between generic sectors and null string currents via a 
slight rescale of drifts and interactions. 

When it comes to mesoscopic levels of description (Sec. III\,B), also distinctive 
features appeared at low temperature regimes. In spite of the residual entropy in
most of the studied sectors (stemming from their highly degenerate ground states), 
under repulsive couplings there is a substantial increase of both correlation lengths 
and structure factors at characteristic wavenumbers as temperature is lowered 
(Figs.\,\ref{SF}a and \ref{SF}b). We interpret these modes as being selected by 
thermal fluctuations from the ground state manifold thus giving rise, we suggest, to 
an order-by-disorder scenario \cite{Villain}. For attractive interactions however, this 
latter can not be inferred from our simulations since, in line with the ground states
degeneracies, both $\xi$ and $S(q)$ do not grow any further; a  situation which 
resembles that of the standard DLG $(\rho \ne 1/2)$, and null strings $({\cal L}/L \ne
\frac{1}{k+1})$ under repulsive couplings (Fig.\,\ref{SFnull}). Finally, both large 
drive regimes and non-interacting sectors are governed by a simple product 
measure, but due to the $k$-mer size their correlation functions may show nontrivial 
oscillations and large correlation lengths (Figs.\,\ref{SF}a, \ref{SF}b, and \ref{SF}c)
\cite{Gupta}.

It is natural to ask whether the above numerical findings could be approached 
theoretically. At the microscopic level of the master equation \cite{Kampen}, the 
formal analogies between this latter and the Schr\"odinger equation describing the
evolution of associated quantum spin chains have proven useful in the analysis
of several nonequilibrium processes \cite{Schutz,me}. In fact, for vanishing drives
and interactions these reconstructing systems have been studied in terms of 
spin-$\frac{1}{2}$ Heisenberg ferromagnets \cite{Menon}, but for ${\cal E},\,
\beta J \ne 0$ the evolution operators are neither familiar nor simple to analyze. On 
the other hand, already at the mean field level it is not clear how to proceed with an 
exponential number of conservation laws such as the IS's discussed throughout.

Other issues not covered here that would be worth pursuing concern the phase
ordering dynamics of periodic sectors under repulsive couplings where, as we have
pointed out, stationary correlation lengths can get very large at low temperature 
regimes. It would be intersting to determine whether the dynamic exponents 
characterizing the large time growth of $\xi$ actually depend on the subspaces where
the evolution takes place. There is also the question about tagged particle diffusion
either with or without driving fields. For $\beta J = 0$ and ${\cal E} \ne 0$ it is known
that in the non-reconstructing case the root-mean-square displacement of a tagged 
particle around its mean position grows asymptotically in time as $\sim t^{1/2}$ 
(as usual), but if the bias is zero it grows anomalously slow as $\sim t^{1/4}\,$ 
\cite{Majumdar}. In the reconstructing situation, where such caging effect might 
depend on the particular distribution of fragments of the sector  considered, these 
issues remain quite open. The question also extends to the coarsening regime 
$- \beta J \gg 1$ for which other displacement laws might emerge depending on
whether or not the motion is driven \cite{Godreche}.

%____________________________
\section*{Acknowledgments}
%____________________________

Support of CONICET and ANPCyT, Argentina under Grants No. PIP 1691 and
No. PICT 1426 is acknowledged.

%____________ References ___________________

%______________________________________________

\end{document}